\documentclass[letterpaper,10 pt,conference]{IEEEtran}
\IEEEoverridecommandlockouts
\IEEEsettopmargin{t}{72pt}
\usepackage{amsmath,amssymb,amsfonts}
\usepackage{graphicx}
\usepackage{textcomp}
\usepackage{xcolor}
\usepackage{lipsum}
\usepackage{svg}
\usepackage{comment}
\def\BibTeX{{\rm B\kern-.05em{\sc i\kern-.025em b}\kern-.08em
    T\kern-.1667em\lower.7ex\hbox{E}\kern-.125emX}}

\usepackage{comment}

\usepackage{nohyperref}

\definecolor{matplotlib0}{HTML}{1f77b4}
\definecolor{matplotlib1}{HTML}{d62728}
\definecolor{matplotlib2}{HTML}{2ca02c}
\definecolor{matplotlib3}{HTML}{ff7f0e}
\definecolor{matplotlib4}{HTML}{9467bd}
\definecolor{matplotlib5}{HTML}{8c564b}
\definecolor{matplotlib6}{HTML}{e377c2}
\definecolor{matplotlib7}{HTML}{7f7f7f}
\definecolor{matplotlib8}{HTML}{bcbd22}
\definecolor{matplotlib9}{HTML}{17becf}

\usepackage{mathtools} 

\usepackage{booktabs}
\usepackage{multirow}
\usepackage{colortbl}
\usepackage{tablefootnote}
\usepackage{threeparttable}

\usepackage[acronym, style=super, nonumberlist]{glossaries}

\usepackage{pgfplots}
\definecolor{color0}{rgb}{0.12156862745098,0.466666666666667,0.705882352941177} 
\definecolor{color1}{rgb}{1,0.498039215686275,0.0549019607843137}
\definecolor{color2}{rgb}{0.172549019607843,0.627450980392157,0.172549019607843} 
\definecolor{color3}{rgb}{0.83921568627451,0.152941176470588,0.156862745098039} 
\definecolor{color4}{rgb}{0.580392156862745,0.403921568627451,0.741176470588235}
\definecolor{colorblue}{rgb}{0.12156862745098,0.466666666666667,0.705882352941177} 
\definecolor{colorgreen}{rgb}{0.172549019607843,0.627450980392157,0.172549019607843} 
\definecolor{colorred}{rgb}{0.83921568627451,0.152941176470588,0.156862745098039} 
\definecolor{colorblack}{rgb}{0,0,0} 
\definecolor{colororange}{rgb}{1,0.56,0} 
\usepgfplotslibrary{fillbetween}
\usepgfplotslibrary{colormaps}
\pgfplotsset{compat=1.16}

\pgfplotscreateplotcyclelist{matplotlib}{
  {matplotlib0},
  {matplotlib1},
  {matplotlib2},
  {matplotlib3},
  {matplotlib4},
  {matplotlib5},
  {matplotlib6},
  {matplotlib7},
  {matplotlib8},
  {matplotlib9}
}

\pgfplotsset{every axis/.append style={
    cycle list name=matplotlib,
}}

\usepackage{listings}

\definecolor{code_default}{HTML}{000000}
\definecolor{code_keyword}{HTML}{AC4142}
\definecolor{code_identifier}{HTML}{D28445}

\lstdefinelanguage{RISCV}{
  sensitive=false,
  morecomment=[l]{//},
  alsoletter={.},
  morekeywords=[1]{
    lp.setup, mv, lw, p.lw, sw, p.sw, pv.sdotsp.b, pv.shuffle2.b, p.subNR, p.addNR
  },
  morekeywords=[2]{
    zero, ra, sp, gp, tp, t0, t1, t2, t3, t4, t5, t6, s0, s1, a0, a1, a2, a3, a4, a5, a6, a7, a8, a9, a10, a11,
  },
  morestring=[b]",
  morestring=[b]',
}[strings, comments, keywords]

\lstdefinestyle{RISCV_STYLE}{
  language=RISCV,
  numbers=none,
  basicstyle=\scriptsize\ttfamily\color{code_default},
  keywordstyle=[1]\color{matplotlib0},
  keywordstyle=[2]\color{matplotlib1},
  float,
  captionpos=b,
  belowskip=-0.5cm
}

\usepackage{algorithm}
\usepackage{algpseudocode}
\usepackage{float}
\newfloat{algorithm}{t}{top}

\newacronym{simd}{SIMD}{Single Instruction, Multiple Data}
\newacronym{elu}{ELU}{Exponential Linear Unit}
\newacronym{relu}{ReLU}{Rectified Linear Unit}
\newacronym{rpr}{RPR}{Random Partition Relaxation}
\newacronym{mac}{MAC}{Multiply Accumulate}
\newacronym{dma}{DMA}{Direct Memory Access}
\newacronym{bmi}{BMI}{Brain--Machine Interface}
\newacronym{bci}{BCI}{Brain--Computer Interface}
\newacronym{smr}{SMR}{Sensory Motor Rythms}
\newacronym{eeg}{EEG}{Electroencephalography}
\newacronym{svm}{SVM}{Support Vector Machine}
\newacronym{svd}{SVD}{Singular Value Decomposition}
\newacronym{evd}{EVD}{Eigendecomposition}
\newacronym{iir}{IIR}{Infinite Impulse Response}
\newacronym{fir}{FIR}{Finite Impulse Response}
\newacronym{fc}{FC}{Fabric Controller}
\newacronym{nn}{NN}{Neural Network}
\newacronym{mrc}{MRC}{Multiscale Riemannian Classifier}
\newacronym{flop}{FLOP}{Floating Point Operation}
\newacronym{sos}{SOS}{Second-Order Section}
\newacronym{ipc}{IPC}{Instructions per Cycle}
\newacronym{tcdm}{TCDM}{Tightly Coupled Data Memory}
\newacronym{fpu}{FPU}{Floating Point Unit}
\newacronym{fma}{FMA}{Fused Multiply Add}
\newacronym{alu}{ALU}{Arithmetic Logic Unit}
\newacronym{dsp}{DSP}{Digital Signal Processing}
\newacronym{gpu}{GPU}{Graphics Processing Unit}
\newacronym{soc}{SoC}{System-on-Chip}
\newacronym{mi}{MI}{Motor-Imagery}
\newacronym{csp}{CSP}{Commmon Spatial Patterns}
\newacronym{fbcsp}{FBCSP}{Filter-Bank \acrlong{csp}}
\newacronym{pulp}{PULP}{parallel ultra-low power}
\newacronym{soa}{SoA}{state-of-the-art}
\newacronym{bn}{BN}{Batch Normalization}
\newacronym{isa}{ISA}{Instruction Set Architecture}
\newacronym{ecg}{ECG}{Electrocardiogram}
\newacronym{mcu}{MCU}{microcontroller}
\newacronym{rnn}{RNN}{recurrent neural network}
\newacronym{cnn}{CNN}{convolutional neural network}
\newacronym{tcn}{TCN}{temporal convolutional network}
\newacronym{emu}{EMU}{epilepsy monitoring unit}

\def\ps@IEEEtitlepagestyle{%
  \def\@oddfoot{\mycopyrightnotice}%
  \def\@oddhead{\hbox{}\@IEEEheaderstyle\leftmark\hfil\thepage}\relax
  \def\@evenhead{\@IEEEheaderstyle\thepage\hfil\leftmark\hbox{}}\relax
  \def\@evenfoot{}%
}
\def\mycopyrightnotice{%
  \begin{minipage}{\textwidth}
  \centering \scriptsize
  \copyright 2022 IEEE.  Personal use of this material is permitted.  Permission from IEEE must be obtained for all other uses, in any current or future media, including reprinting/republishing this material for advertising or promotional purposes, creating new collective works, for resale or redistribution to servers or lists, or reuse of any copyrighted component of this work in other works.
  \end{minipage}
}
\makeatother

\begin{document}
\thispagestyle{empty}
\pagestyle{empty}

\title{
Energy-Efficient Tree-Based EEG Artifact Detection 
}

 \author{\IEEEauthorblockN{
    Thorir Mar Ingolfsson\IEEEauthorrefmark{1},
    Andrea Cossettini\IEEEauthorrefmark{1},
    Simone Benatti\IEEEauthorrefmark{2}\IEEEauthorrefmark{3},
    Luca Benini\IEEEauthorrefmark{1}\IEEEauthorrefmark{2}}
     

    \IEEEauthorblockA{\IEEEauthorrefmark{1}Integrated Systems Laboratory, ETH Z{\"u}rich, Z{\"u}rich, Switzerland}
    \IEEEauthorblockA{\IEEEauthorrefmark{2}DEI, University of Bologna, Bologna, Italy}
    \IEEEauthorblockA{\IEEEauthorrefmark{3}DIEF, University of Modena and Reggio Emilia, Reggio Emilia, Italy}

    \thanks{Corresponding email: \{thoriri\}@iis.ee.ethz.ch}
    
    }

\maketitle

\begin{abstract}
In the context of epilepsy monitoring, EEG artifacts are often mistaken for seizures due to their morphological similarity in both amplitude and frequency, making seizure detection systems susceptible to higher false alarm rates. In this work we present the implementation of an artifact detection algorithm based on a minimal number of EEG channels on a parallel ultra-low-power (PULP) embedded platform. The analyses are based on the TUH EEG Artifact Corpus dataset and focus on the temporal electrodes. First, we extract optimal feature models in the frequency domain using an automated machine learning framework, achieving a 93.95\% accuracy, with a 0.838 F1 score for a 4 temporal EEG channel setup.
The achieved accuracy levels surpass state-of-the-art by nearly 20\%. Then, these algorithms are parallelized and optimized for a PULP platform, achieving a 5.21$\times$ improvement of energy-efficient compared to state-of-the-art low-power implementations of artifact detection frameworks. Combining this model with a low-power seizure detection algorithm would allow for 300h of continuous monitoring on a 300 mAh battery in a wearable form factor and power budget. These results pave the way for implementing affordable, wearable, long-term epilepsy monitoring solutions with low false-positive rates and high sensitivity, meeting both patients' and caregivers' requirements.
\newline
\indent \textit{Clinical relevance}— The proposed EEG artifact detection framework can be employed on wearable EEG recording devices, in combination with EEG-based epilepsy detection algorithms, for improved robustness in epileptic seizure detection scenarios. 
\end{abstract}

\begin{IEEEkeywords}
healthcare, time series classification, smart edge computing, machine learning, deep learning
\end{IEEEkeywords}

\section{Introduction}\label{ch:introduction}

\gls{eeg} is the golden standard medical diagnostic tool to analyze brain activity, with applications that include developing brain-machine interfaces or monitoring epilepsy \cite{world2019epilepsy}.
Conventional EEG systems are, however, bulky and uncomfortable, causing patients to commonly refer perceived stigmatization. Furthermore, the long wires used to connect multiple electrodes are a significant cause of motion artifacts on the EEG traces \cite{tatum2011artifact}. Thus, patients and caregivers firmly push for wearable EEG solutions based on a minimal number of electrodes, allowing for continuous monitoring during the day in an inconspicuous way~\cite{bruno2020seizure}.

Several groups are involved in the development of such minimal EEG systems that can take the form of glasses, behind-the-ear, or even in-ear solutions~\cite{asif2020epileptic,pham2020wake,sopic2018glass,guermandi2018wearable,bvba_byteflies_nodate,kaveh2020wireless,gu2018comparison,do2014wireless}.
However, the collection of \gls{eeg} signals is not trivial, and signals can become contaminated with artifacts even in the absence of long cables. In fact, artifacts can be caused by external sources (e.g., electrode movement, external noise) or by physiological factors  (e.g., chewing, muscle movements, electromyographic or electrooculogram signals coupled to \gls{eeg}) that are not of cerebral origin. In the context of epilepsy monitoring, these \gls{eeg} artifacts are often mistaken for seizures due to their morphological similarity in both amplitude and frequency~\cite{9353647}, hampering the epilepsy detection algorithms. It is therefore of paramount importance that \gls{eeg} artifacts are properly detected so as not to induce false alarms in epilepsy detection systems. After detecting an artifact successfully, the epilepsy detection system could then intelligently decide whether to adaptively correct the \gls{eeg} signal (using methods such as Artifact Subspace Reconstruction~\cite{8768041}), or to extract features that can help the system classify upcoming seizures or disregard the data. To this end, there is a need for estimations of the generic presence of artifacts (binary classification, BC), if not a per-channel estimation of the presence of artifacts (multi-label classification, MC), or even better, an estimation of the specific source cause of artifacts on each channel (multi-class multi-output classification, MMC).

Many different artifact detection/rejection algorithms have been proposed~\cite{mognon2011adjust,lawhern2012detection,8584791,9137056}, relying on both supervised and unsupervised methods. In~\cite{8584791}, an energy-efficient convolutional neural network (CNN) is used to process raw \gls{eeg} signals without the need for feature extraction, achieving around 74\% accuracy on an in-house \gls{eeg} artifact dataset. In~\cite{roy2019machine}, which focuses on the Temple University Hospital (TUH) EEG Artifact Corpus (TUAR)~\cite{9353647} dataset, a $71.43\%$ accuracy score and a $0.8021$ weighted F1 score was obtained, utilizing a \emph{LDA} model and frequency features. In another work~\cite{qendro2021high}, also focused on the TUAR dataset, a deep learning framework was used, achieving a $0.838$ F1 score. Finally, also in~\cite{Kim2021} the focus was on the TUAR dataset, and deep learning models allowed to classify artifacts with a $75\%$ accuracy.

However, such approaches present significant limitations in the context of the deployment on minimal, wearable, low-power epilepsy monitoring devices.
In fact, most of these methods and algorithms have not been designed with wearable long-term systems in mind. For example, the energy-efficient CNN introduced in~\cite{8584791} has an energy consumption of around 31 mJ per inference, which is not suitable for long-term usage. The Low-Power LSTM Processor introduced in~\cite{9137056} has an energy consumption of around 21 {\textmu}J per inference, which is a step in the right direction. Further optimizations are, however, needed, also to increase the accuracy of the detection.

Within this context, we present a framework to automatically detect EEG artifacts, designed to aid and support more robust epilepsy detection systems~\cite{9644949} for wearable unobtrusive epilepsy detection with a reduced montage. Specifically, the main contributions are the following:
\begin{itemize}
    \item Optimal model selection using a publicly available automated machine learning framework;
    \item A comprehensive evaluation of models on 15 datasets extracted from the TUH EEG Artifact Corpus;
    \item Achieved state-of-the-art $\approx$94\% artifact detection accuracy considering a 4 temporal channel EEG setup;
    \item Implementation and performance optimization of the above framework on real PULP chip target, namely Mr. Wolf~\cite{pullini2019mr}, designed in 40nm FDSOI technology and widely used in ultra-low-power wearable nodes for biomedical applications~\cite{kartsch2019biowolf}, achieving state-of-the-art $\approx$4 {\textmu}J per inference.
\end{itemize}
The experimental results demonstrate that a high-accuracy framework for continuous automatic artifact detection can be designed for wearable long-lifetime systems with a low electrodes count.

\section{Background}\label{ch:background}

\subsection{Artifact Dataset}\label{ch:background:dataset}
The TUH EEG Corpus is the world’s most extensive open-source corpus of EEG data~\cite{9353647}. A subset of the corpus is the Temple University Artifact Corpus (TUAR), containing 310 annotated EEG files from 213 patients with annotations of every artifact. The dataset includes labels of five different individual artifacts, seven combinations of two different artifacts, and background EEG (no artifact), with a total of 13 different labels. Five different sampling frequencies are used in the dataset, with around $71.51\%$ of the data sampled at 250 Hz.

\subsection{Related Work}\label{ch:background:related}
Not much prior work has been done on the TUAR dataset concerning artifact detection. One of the first works~\cite{roy2019machine} was set up as a benchmark for further explorations on the dataset: it dates back to 2019 and used an earlier release of the TUAR dataset, where not every artifact had been fully annotated. Furthermore, while all the data have been considered, a clear indication of how the different sampling frequencies are processed was missing. Fast Fourier Transform (FFT) was utilized for feature pre-processing, and the algorithms considered are the following: \emph{AdaBoost, Gaussian Naive Bayes, k-nearest neighbors, Linear discriminant analysis (LDA), multilayer perceptron, Random Forest, stochastic gradient descent Classifier, XGBoost}. The primary outcome of the work was that the best performing algorithm was the \emph{LDA} with a $71.43\%$ accuracy score and a $0.8021$ weighted F1 score.

In~\cite{9413712}, a 1D-CNN architecture takes in the raw EEG data with no additional feature pre-processing. The primary purpose of the work was not to classify artifacts with as much accuracy as possible but instead to design a framework that works when electrodes are missing or shifted. As such, the baseline accuracy $71.1\%$ does not improve upon the results offered by \emph{LDA}.

A deep learning framework for an early exit upon artifact detection was proposed by~\cite{qendro2021high}. Since the dataset has multiple sampling frequencies, the authors first re-sample the data that do not adhere to the 250 Hz sampling frequency. Then, they band-pass filter the data between 0.3 and 40 Hz and apply a notch filter at 60Hz. Windows of 10 seconds length are considered, and each channel is treated independently. The dataset was then split into 80-10-10 for train-validation-test, patient-independent between the datasets. The reported results are $0.838$ F1 score, $0.853$ precision, and $0.829$ recall.

The most extensive research on this dataset for artifact detection is done in~\cite{Kim2021}. The same procedure as of~\cite{qendro2021high} is followed, with the re-sampling of the data to 250 Hz. Then, three different deep learning methods and an ensemble of their outputs are tested. The best accuracy score for the binary case was obtained with a CNN network achieving $75\%$ accuracy, and the best result for the ensemble resulted in a $67.59\%$ accuracy.


\section{Materials \& Methods}\label{ch:background}
\begin{figure}[t]
  \centering
   \includegraphics[width=0.9\linewidth]{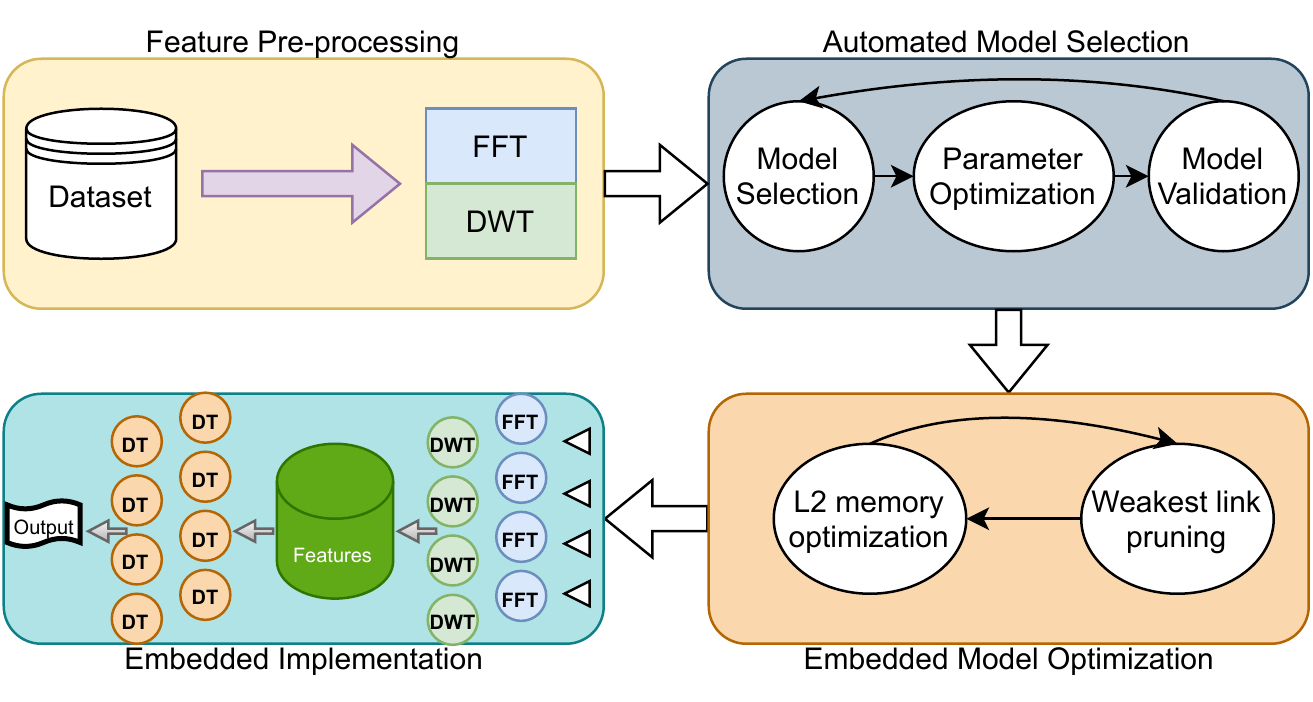}
  \vspace{-0.2cm}
  \caption{Schematic representation of the workflow. Data is pre-processed using DWT and FFT, and model selection is made with TPOT. Embedded optimization and pruning are then done before classification is performed with an ensemble of decision trees.}
  \label{fig:workflow_sketch}
  \vspace{-0.2cm}
\end{figure}

In this section, we present the main steps followed in this work, starting from the extraction and re-sampling of the dataset, followed by the pre-processing and model selection, and concluding with the implementation on the PULP embedded platform. 
Fig.~\ref{fig:workflow_sketch} shows a sketch of the workflow.

\subsection{Dataset extraction}

The TUAR has 22 channels separately annotated with 13 different labels (12 different artifacts labels, 1 non-artifact label). Starting from this dataset, we consider three different types of classification approaches: BC, MC, MMC. In the following, we indicate as ${\Delta}T$ a generic time window, and as $\Theta\left(\Delta T \right)$ the label assigned to that time window.

\textit{Binary Classification (\textbf{BC})}.
For a time window ${\Delta}T$, if there is an artifact on any one of the channels, we label that window as an artifact ($\Theta\left(\Delta T \right)=1$). Differently, if no channel was initially labeled with an artifact in that window, the window gets labeled as standard background EEG ($\Theta\left(\Delta T \right)=0$).

\textit{Multilabel Classification (\textbf{MC})}.
In this scenario, each channel is analyzed independently. Therefore, we have an independent binary classification on each channel. If the $i$-th channel was initially labeled with any artifact label, we assign a label to that window for that channel as $\Theta_i \left(\Delta T \right)=1$. Similarly, if channel $i$ does not have an artifact in the considered time window, it is labeled as normal background EEG: $\Theta_i \left(\Delta T \right)=0$.

\textit{Multiclass-Multioutput Classification (\textbf{MMC})}.
In this scenario, we expand the MC, discriminating between the specific cause of the artifact (12 alternative possibilities) for each label of each channel. Therefore, if the $i$-th channel was originally labeled with the artifact type $k$ (with $k \in \left[1,12 \right]$), we assign a label to that window for that channel as $\Theta_i \left(\Delta T \right)=k$. Similarly, if channel $i$ does not have an artifact in the considered time window, it is labeled as normal background EEG: $\Theta_i \left(\Delta T \right)=0$.

The TUAR dataset also has 5 different sampling frequencies (250 Hz, 256 Hz, 400 Hz, 512 Hz, and 1000 Hz). Since the features extraction requires having a model with a constant sampling frequency, we also extract 5 different datasets depending on the sampling frequencies.
\begin{itemize}
    \item 250 Hz \textbf{(A)}
    \item 250 + 1000 Hz \textbf{(B)}
    \item 256Hz \textbf{(C)}
    \item 256 + 512 Hz \textbf{(D)}
    \item All frequencies \textbf{(E)}
\end{itemize}
Dataset \textbf{(A)} contains only the data of the original dataset that was sampled at 250 Hz. Datasets \textbf{B} and \textbf{C} are generated by decimating the higher frequency to match the lower one. Dataset \textbf{E} is generated by re-sampling the data to match a 250 Hz sampling frequency, using a one-dimensional linear interpolation.

To summarize, we extract 15 different datasets from the TUAR data repository, considering 5 different frequency configurations and 3 different labeling approaches.

\subsection{Pre-processing and feature extraction}
We use Discrete Wavelet Transform (DWT) and FFT for feature extraction on each dataset. We consider a 1 second time window as longer windows did not result in better classification accuracy of artifacts; furthermore, longer windows introduce a longer latency of artifact detection.

FFT is a fast way of calculating the frequency representation of time-domain signal values. However, the time information of the signal is lost after the transformation. FFT features have been shown as a good feature for artifact detection~\cite{ISLAM2016287}, and we use the FFT to calculate the energy of the high-frequency parts of the signal (frequencies above 80 Hz). 

DWT is a  signal decomposition technique for time-frequency analyses~\cite{mallat1989theory} that has been reported to work well for seizure detection \cite{9644949}. DWT offers a good trade-off between signal-to-noise ratio and performance \cite{benatti2014towards}, and it allows to obtain both frequency and temporal features of a signal by applying a cascade series \textit{Mother Wavelets} filters. 

We use FFT and DWT with a 1\,s window length. The multi-core platform (Sec.~\ref{sect:mrwolf}) allows the process of computations in a fast, energy-efficient, and accurate manner, splitting the computation of the FFT and DWT using 4 cores for each. Selection of variables and decomposition is performed following the same approach as in~\cite{7833731}. For the DWT, we use the Haar wavelet, and once the detail coefficients for each level are computed, we calculate the energy to be used as a feature. The Haar wavelet was chosen through empirical exploration. Even though the FFT and DWT features can be used alone and give good metrics, relying on a combination of FFT and DWT features yielded the best results.
\begin{figure}[h!]
  \centering
  \vspace{-0.3cm}
  \includegraphics[width=0.58\linewidth, clip, trim={0 0 0 0}]{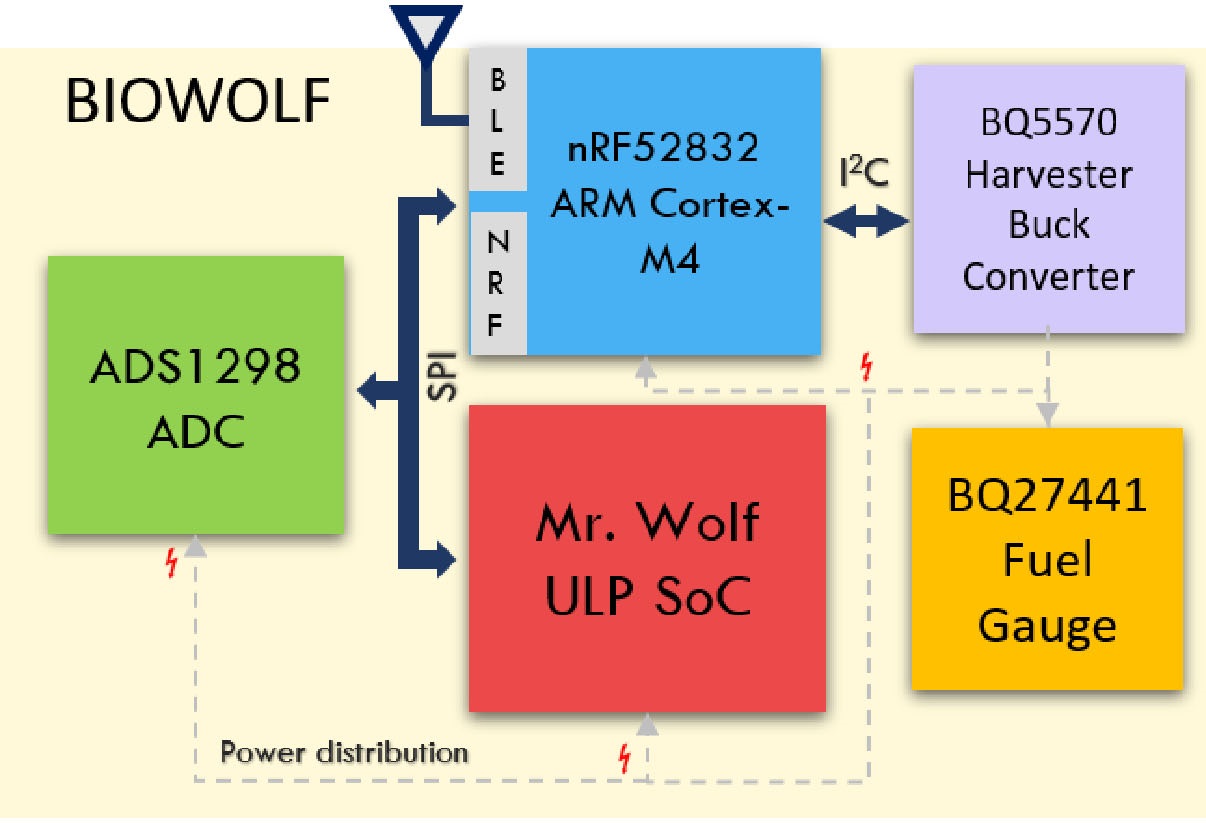}
  \includegraphics[width=0.29\linewidth, clip, trim={0 0 0 0}]{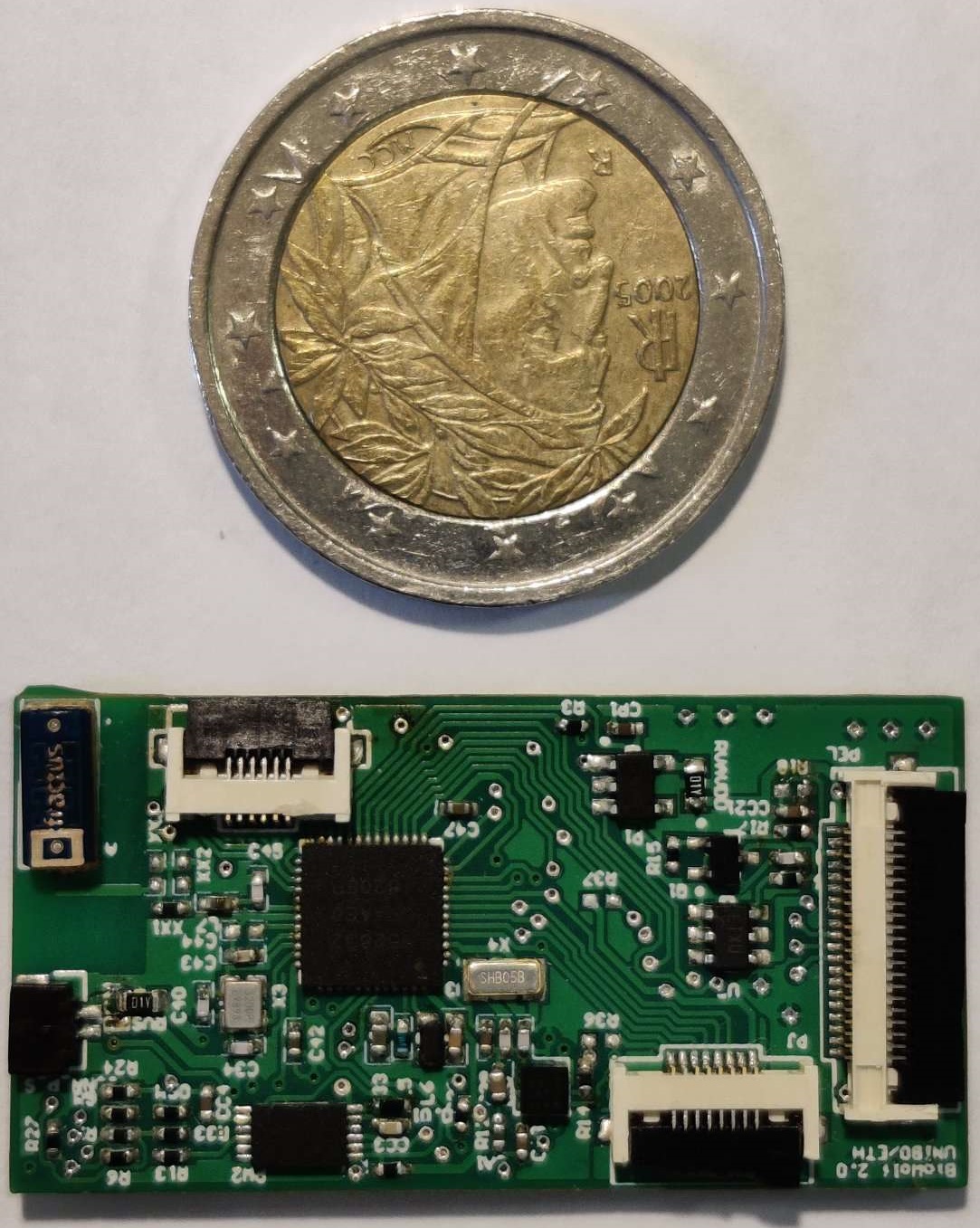}
  \vspace{-0.2cm}
  \caption{Left: main building blocks of the BioWolf~\cite{kartsch2019biowolf} embedded platform. Right: wearable form factor of BioWolf, compared to a 2 euro coin.}
  \label{fig:MrWolf}
  \vspace{-0.2cm}
\end{figure}

\subsection{Model Selection}
Following the feature extraction based on FFT and DWT, for model selection and optimization, we utilize the Tree-based Pipeline Optimization Tool (TPOT)~\cite{le2020scaling}, which is an automated machine learning system that takes in features and labels and uses genetic programming to output the best model with cross-validated classification accuracy. Using TPOT allows for a comprehensive search over a wide field of different machine learning models. Among the different available AutoML frameworks~\cite{agtabular,H2OAutoML20,feurer-neurips15a}, we relied on TPOT because of its simplicity and short development time. The reported methodology can be easily adapted and extended to the other frameworks.

\subsection{Performance Metrics}
We evaluate the models according to the classification accuracy, the ratio between correctly classified trials, and the total number of trials in the test set. 
Since the TUAR dataset we use in this paper is imbalanced, we also consider the F1 score, which is defined as:

\begin{equation}
F_1 = 2 \frac{\textrm{Precision}\cdot\textrm{Recall}}{\textrm{Precision}+\textrm{Recall}}
\end{equation}

For the \textbf{MMC} case, we report weighted F1 scores considering each class's support.
\begin{table*}[ht!]
\renewcommand{\arraystretch}{1}
  \centering
  \caption{Optimized algorithm output from TPOT}\label{tab:results:summary}
  \vspace{-.2cm}
  {
\begin{tabular}{@{}lrrrrrrrrrrrrrrr@{}}
\toprule
\textit{Sampling rate}& \multicolumn{3}{c}{250 Hz} & \multicolumn{3}{c}{250+1000 Hz} & \multicolumn{3}{c}{256 Hz} & \multicolumn{3}{c}{256+512 Hz} & \multicolumn{3}{c}{All}\\
\cmidrule(lr){2-4} \cmidrule(lr){5-7} \cmidrule(lr){8-10} \cmidrule(l){11-13} \cmidrule(l){14-16} 
\textit{Dataset} & \multicolumn{1}{r}{BC}   & \multicolumn{1}{r}{MC}    & \multicolumn{1}{r}{MMC}   & \multicolumn{1}{r}{BC}   & \multicolumn{1}{r}{MC}    & \multicolumn{1}{r}{MMC}    & \multicolumn{1}{r}{BC}   & \multicolumn{1}{r}{MC}    & \multicolumn{1}{r}{MMC}    & \multicolumn{1}{r}{BC}   & \multicolumn{1}{r}{MC}    & \multicolumn{1}{r}{MMC}   & \multicolumn{1}{r}{BC}   & \multicolumn{1}{r}{MC}    & \multicolumn{1}{r}{MMC} \\ 
\midrule

\hspace{1mm}Accuracy [\%]          & 93.95   & 90.05   & 89.23   & 93.43 & 90.14   & 89.25   & 87.34   & 86.75   & 86.92 & 86.88 & 86.52 & 85.69 & 92.23 & 90.72 & 87.94 \\ 
\hspace{1mm}F1 score         & 0.838   & 0.600   & 0.867   & 0.825 & 0.608   & 0.868   & 0.644  & 0.607  & 0.830 & 0.671 & 0.647 & 0.826 & 0.798 & 0.749 & 0.853\\
\bottomrule
\end{tabular}
  }
  \vspace{-0.5cm}
\end{table*}

\subsection{Embedded platform}\label{sect:mrwolf}

Following the approach of \cite{9644949}, we implement the artifact detection framework on the BioWolf wearable ExG device~\cite{kartsch2019biowolf} (Fig. \ref{fig:MrWolf}). BioWolf is a sub-10-mW 8-channels brain-computer interface platform equipped with the PULP Mr. Wolf multi-core processor and offering wireless Bluetooth low-energy (BLE) connectivity. The BioWolf system is based on a multi-core fully programmable processor, combining the versatility of a programmable system with the efficiency of a computational 8-core cluster with two shared floating-point units. Mr. Wolf outperforms the performance achievable with commercial single-core low-power processors with a comparable power budget by at least one order of magnitude. Mr. Wolf has an L2 memory of 512 kB and a shared L1 memory of 64 kB in terms of memory footprint.

\subsection{Embedded implementation}
\subsubsection{Feature Extraction Implementation}
The FFT implementation on Mr. Wolf exploits the conjugate symmetry of real-valued FFT to optimize the computation. A complex FFT is computed over half of the signal, and the real-valued part is then extracted from the output. Therefore, the computation is significantly reduced compared to doing a full real-valued FFT on all the signal. Furthermore, the FFT also uses a mixed-radix complex FFT approach to save processor computation time. As for the DWT implementation, it iteratively passes through low-pass and high-pass filters.
\subsubsection{Decision Tree Implementation}
The optimized temporal models generated by TPOT are mainly Extra Trees, i.e., an ensemble of Decision Trees~\cite{Rokach2005} (DT). The obtained DTs are pretty large, with millions of threshold values, not fitting on an embedded platform. Therefore, we applied pruning on the DTs with a Minimal cost-complexity pruning (MCCP) algorithm~\cite{Breiman1983ClassificationAR}. While MCCP is most often used to prune DTs to avoid overfitting, in this case, we use it to prune the DTs down such that the tree will fit on the L2 memory of the Mr. Wolf multicore processor. We, therefore, take the best-performing model that TPOT outputs and prune it such that the total number of nodes in the tree fits on the 512 kB L2 memory. 

During inference, we start from the tree's root node and look in the current feature array to see which feature we must compare with the current node threshold value. Depending on the outcome of the comparison between the threshold value and the feature value, we either go into the tree's left or right branches.
The behavior of a decision tree can then be implemented using four arrays:
\begin{itemize}
    \item Current feature array
    \item Threshold value array
    \item Left child node index array 
    \item Right child node index array
\end{itemize}
Usually, a decision tree implementation needs another array: a value array that corresponds to the final classification output of the tree. Our four array approach is, therefore, an optimized representation of the decision tree since we inject the corresponding class value prediction into the right child node at the current index for leaf nodes. By doing so, we compress the required memory footprint to implement a decision tree by $10-30\%$ (depending on how the value array is implemented).

We represent the threshold values with 32 bits. Since the feature array is relatively small (5 features per channel), we can represent it with 8 bits, while the index of the left and right child nodes can be represented with 16 bits. In total, each node in the decision tree requires 9 bytes. These arrays are then stored on the L2 memory of Mr. Wolf.

We do not use double-buffered direct memory transfer to transfer tree contents from L2 to L1. This decision is based on the sparse inference property most decision trees have. The maximum number of nodes in a decision tree depends on the maximum depth of the tree, with the number of nodes (binary decision trees) growing exponentially with regards to the depth ($2^{d+1} - 1$). However, only a tiny percentage of these nodes are traversed during inference.
As an example, we can consider a decision tree of depth 20. Such tree has a theoretical maximum number of nodes of $2\,097\,151$. In practice, such a decision tree with a maximum depth of 20 would have far fewer nodes, as some branches are terminated early. During inference of this example DT, we would at most traverse through 20 nodes, which is only around $0.00095\%$ of the total nodes. Consequently, doing double-buffered direct memory access would require the microcontroller to move $2\,097\,151$ nodes (around 18.9 MB) from the L2, while the inference only requires 20 nodes (around 180B). This explains why blind direct memory access (DMA) of all nodes to L1 is an unnecessary and energy-wasting solution.

 \subsection{Power Measurements}
We operate the cluster of Mr. Wolf at 100 MHz (frequency of maximal energy efficiency~\cite{pullini2019mr}). A Keysight N6705C power analyzer, with 0.2048\,ms sampling interval, is used to perform the power measurements.

\begin{figure*}[t]
\centering

\begin{minipage}{0.49\textwidth}

  \resizebox{\textwidth}{!}{%
    \centering

\begin{tikzpicture}
\begin{axis}[
name=ax1,
height=\axisdefaultheight,
width=\textwidth,
tick pos=left,
legend cell align={left},
legend style={
        at={(0.99, 0.01)},
        anchor=south east,
        draw=none,
        font={\scriptsize}
      },
x grid style={black!20},
xlabel={Size of model [kB]},
xmajorgrids,
xlabel near ticks,
xticklabel style={
        /pgf/number format/fixed,
        /pgf/number format/precision=5
},
scaled x ticks=false,
xmin=0, xmax=12000,
y grid style={black!20},
ymajorgrids,
ylabel near ticks,
ylabel={TUAR Test Accuracy [\%]},
ylabel near ticks,
ymin=85, ymax=95,
]

\addplot [semithick, colorblue, dotted, mark=*, mark size=1.5, mark options={solid}]
table[x = x, y = y] {%
250.dat
};
\addlegendentry{BC}
\addplot [semithick, colorblack, dotted, mark=*, mark size=1.5, mark options={solid}]
table[x = x, y = y] {%
250_MC.dat
};
\addlegendentry{MC}
\addplot [semithick, colorred, dotted, mark=*, mark size=1.5, mark options={solid}]
table[x = x, y = y] {%
250_MMC.dat
};
\addlegendentry{MMC}
\addplot [very thick, colororange]
table[x = x, y = y] {%
redline.dat
};

 \node [draw=none,anchor=south west, colororange] at  (600, 94) {\scriptsize\shortstack[l]{
 L2 Memory Capacity
}};

\draw[-latex,very thick,draw=colorblue] (axis cs:9500,93) -- (axis cs:11150,93.8);
 \node [draw=none,anchor=north west,colorblue] at  (7500, 92.8) {\scriptsize\shortstack[l]{
 Optimal model 93.95\%
}};
 \node [draw=none,anchor=south east,colorblack] at  (11500, 90.9) {\scriptsize\shortstack[l]{
 334.1 MB
}};
\draw[-latex,very thick,draw=colorblack] (axis cs:9500,90.9) -- (axis cs:11600,90.9);
 \node [draw=none,anchor=north west,colorblack] at  (7500, 90.8) {\scriptsize\shortstack[l]{
 Optimal model 90.05\%
}};

\draw[-latex,very thick,draw=colorred] (axis cs:9500,88) -- (axis cs:11600,88);
 \node [draw=none,anchor=north west,colorred] at  (7500, 87.8) {\scriptsize\shortstack[l]{
 Optimal model 89.23\%
}};
 \node [draw=none,anchor=south east,colorred] at  (11500, 88) {\scriptsize\shortstack[l]{
 87.1 MB
}};
\addplot[semithick,colorblue, mark=otimes*,mark size=2,mark options={solid}]
table[x = x, y = y] {%
250_opt.dat
};
\end{axis}

\end{tikzpicture}
}
\end{minipage}
\begin{minipage}{0.49\textwidth}

  \resizebox{\textwidth}{!}{%
    \centering

\begin{tikzpicture}
\begin{axis}[
name=ax1,
height=\axisdefaultheight,
width=\textwidth,
legend cell align={left},
legend style={
        at={(0.99, 0.01)},
        anchor=south east,
        draw=none,
        font={\scriptsize}
      },
tick pos=left,
x grid style={black!20},
xlabel={Size of model [kB]},
xmajorgrids,
xlabel near ticks,
xticklabel style={
        /pgf/number format/fixed,
        /pgf/number format/precision=5
},
scaled x ticks=false,
xmin=0, xmax=600,
y grid style={black!20},
ymajorgrids,
ylabel near ticks,
ylabel={TUAR Test Accuracy [\%]},
ylabel near ticks,
ymin=85, ymax=95,
]

\addplot [semithick, colorblue, dotted, mark=*, mark size=1.5, mark options={solid}]
table[x = x, y = y] {%
250.dat
};
\addlegendentry{BC}
\addplot [semithick, colorblack, dotted, mark=*, mark size=1.5, mark options={solid}]
table[x = x, y = y] {%
250_MC.dat
};
\addlegendentry{MC}
\addplot [semithick, colorred, dotted, mark=*, mark size=1.5, mark options={solid}]
table[x = x, y = y] {%
250_MMC.dat
};
\addlegendentry{MMC}
\addplot [very thick, colororange]
table[x = x, y = y] {%
redline.dat
};
\addplot [very thick, colorgreen]
table[x = x, y = y] {%
greenline.dat
};
 \node [draw=none,anchor=north east, colororange] at  (500, 94) {\scriptsize\shortstack[l]{
 L2 Memory Capacity
}};
 \node [draw=none,anchor=north west, colorgreen] at  (64, 94) {\scriptsize\shortstack[l]{
 L1 Memory Capacity
}};
\end{axis}

\end{tikzpicture}
  }
\end{minipage}

\vspace{-6pt}
\caption{\textbf{Left}: Accuracy vs size of Extra Trees model when implemented on the Mr. Wolf microprocessor for all three labelling methods considered. Regions where L1 and L2 are filled are marked with green and orange lines respectively. \textbf{Right}: zoom of the left plot over the 0-600 kB range.}
\label{fig:prune}
\end{figure*}

\section{Experimental Results}\label{ch:results}
\subsection{Classification results}
Table \ref{tab:results:summary} summarizes the accuracy metrics reported for all the datasets and their sampling frequencies considered in this paper, applied to the TUAR data. We consider the TCP average referenced montage~\cite{8257019}, focusing on the 4 temporal channels (F7-T3, T3-T5, F8-T4, T4-T6, according to the 10-20 international system notation).
Comparing the best models which TPOT found for the \textbf{MMC} and all frequencies case, we see that the model outperforms all other models proposed in the literature (which were limited to maximum $75\%$ accuracy and maximum $0.838$ F1 score). These results demonstrate the feasibility of accurate artifact classification based on minimal-montage EEG setups.

\subsection{Embedded Implementation}

After inspecting the optimal DT ensemble that the automated TPOT framework outputs, we optimized it further to fit Mr. Wolf. First, we limit the number of decision trees in the ensemble such that they are a multiple of $8$, to parallelize it on the 8-cores cluster. Then, we use the MCCP algorithm to prune away the weakest links in the decision trees. Figure~\ref{fig:prune} shows the result of such optimization. We perform these limiting and pruning operations for all three cases (BC, MC, MMC) of the 250 Hz dataset. In the \textbf{BC} case, by limiting the number of decision trees to be a multiple of $8$, we can decrease the size of the model very drastically, from $11000$ kB to around $2300$ kB, at the cost of only $1\%$ of accuracy reduction. Further optimization is done using the MCCP algorithm, slowly increasing the complexity parameter that controls how much pruning is done. We can then see that we reach a $92.4\%$ accuracy when the model's size matches the size of the L2 memory (512 kB), which corresponds to a decrease of only $1.55\%$ from the optimal model found by using TPOT. For the \textbf{MC} and \textbf{MMC} cases, we see that we can prune the decision trees more aggressively, with almost no loss in accuracy, until fitting the size of the L1 memory. This low loss of accuracy with pruning validates that embedded implementations of the models with low memory usage can offer similar accuracy levels as state-of-the-art models.

\begin{table}[b]
\vspace{-0.5cm}
\renewcommand{\arraystretch}{1}
  \centering
  \caption{Energy Numbers for implementation on Mr Wolf.}\label{tab:results:summary:wolf}
  \vspace{-.2cm}
  {
    \footnotesize
    \begin{tabular}{@{}lrrrr@{}}
      \toprule
      Dataset & BC & MC & MMC \\
      \midrule
      Time/inference [ms]    &0.18 &0.19 & 0.21 \\
      Power [mW] & 22.41 & 22.43 & 22.44 \\
      Energy/inference [{\textmu}J]   &4.03 &4.26 & 4.71\\
      \bottomrule
    \end{tabular}
  }
\end{table}


Table~\ref{tab:results:summary:wolf} reports the energy numbers of the power measurements performed on Mr. Wolf. Implementing an artifact detecting Extra Trees model on a multi-core edge platform (Mr. Wolf) requires a power envelope of only $\approx$22\,mW with sub-200\,\textmu s processing time, thus with much lower energy requirements than the analog front-end and BLE of the complete platform that enables multi-day functionality~\cite{kartsch2019biowolf}. In combination with previous results for low energy implementation of seizure detection on Mr. Wolf \cite{9644949}, these pave the way to the design of robust and unobtrusive EEG wearable devices.

\section{Conclusion}\label{ch:conclusion}
This work presented the analysis and implementation of an artifact detection framework with minimal EEG setups (4 temporal channels), considering different classification approaches (binary, multi-label, multi-class multi-output). We used a combination of FFT and DWT for signal pre-processing and an automated machine learning framework (TPOT) to search for the optimal model for each scenario. Analyses have been done on the TUAR dataset with 15 sub-datasets extracted, and we achieved state-of-the-art accuracy and F1 scores. Furthermore, we deployed and optimized the algorithms on a PULP platform, achieving minimal energy requirements ($\approx$ 4\,{\textmu}J per inference), outperforming competing commercial devices. These results, combined with robust epilepsy detection models~\cite{9644949}, show that a PULP system is indeed one of the best candidates for future wearable epilepsy monitoring systems based on minimal EEG setups. Future work will focus on integrating the proposed artifact detection model with a robust epilepsy detection framework and on expanding the proposed techniques to include sensor fusion from other data sources, to be realized in a wearable setting of a body area network and on testing the whole system in ambulatory and domestic environments.
\section*{Acknowledgment}
\vspace{-0.1cm}
This project was supported by the Swiss National Science Foundation (Project PEDESITE) under grant agreement 193813.
\bibliographystyle{IEEEtran}
\bibliography{bibliography}

\begin{thebibliography}{10}
\providecommand{\url}[1]{#1}
\csname url@samestyle\endcsname
\providecommand{\newblock}{\relax}
\providecommand{\bibinfo}[2]{#2}
\providecommand{\BIBentrySTDinterwordspacing}{\spaceskip=0pt\relax}
\providecommand{\BIBentryALTinterwordstretchfactor}{4}
\providecommand{\BIBentryALTinterwordspacing}{\spaceskip=\fontdimen2\font plus
\BIBentryALTinterwordstretchfactor\fontdimen3\font minus
  \fontdimen4\font\relax}
\providecommand{\BIBforeignlanguage}[2]{{%
\expandafter\ifx\csname l@#1\endcsname\relax
\typeout{** WARNING: IEEEtran.bst: No hyphenation pattern has been}%
\typeout{** loaded for the language `#1'. Using the pattern for}%
\typeout{** the default language instead.}%
\else
\language=\csname l@#1\endcsname
\fi
#2}}
\providecommand{\BIBdecl}{\relax}
\BIBdecl

\bibitem{world2019epilepsy}
W.~H. Organization \emph{et~al.}, ``Epilepsy: a public health imperative,''
  2019.

\bibitem{tatum2011artifact}
W.~O. Tatum \emph{et~al.}, ``Artifact and recording concepts in eeg,''
  \emph{Journal of clinical neurophysiology}, vol.~28, no.~3, pp. 252--263,
  2011.

\bibitem{bruno2020seizure}
E.~Bruno \emph{et~al.}, ``Seizure detection at home: Do devices on the market
  match the needs of people living with epilepsy and their caregivers?''
  \emph{Epilepsia}, vol.~61, pp. S11--S24, 2020.

\bibitem{asif2020epileptic}
R.~Asif \emph{et~al.}, ``Epileptic seizure detection with a reduced montage: A
  way forward for ambulatory {EEG} devices,'' \emph{IEEE Access}, vol.~8, pp.
  65\,880--65\,890, 2020.

\bibitem{pham2020wake}
N.~Pham \emph{et~al.}, ``{WAKE}: a behind-the-ear wearable system for
  microsleep detection,'' in \emph{Proc. Int. MobiSys}, 2020, pp. 404--418.

\bibitem{sopic2018glass}
D.~Sopic \emph{et~al.}, ``e-glass: A wearable system for real-time detection of
  epileptic seizures,'' in \emph{IEEE ISCAS}, 2018, pp. 1--5.

\bibitem{guermandi2018wearable}
M.~Guermandi \emph{et~al.}, ``A wearable device for minimally-invasive
  behind-the-ear eeg and evoked potentials,'' in \emph{IEEE BioCAS}, 2018, pp.
  1--4.

\bibitem{bvba_byteflies_nodate}
\BIBentryALTinterwordspacing
B.~BVBA, ``\BIBforeignlanguage{en}{Byteflies · {Our} {Solutions}}.''
\BIBentrySTDinterwordspacing

\bibitem{kaveh2020wireless}
R.~Kaveh \emph{et~al.}, ``Wireless user-generic ear {EEG},'' \emph{IEEE
  Transactions on Biomedical Circuits and Systems}, vol.~14, no.~4, pp.
  727--737, 2020.

\bibitem{gu2018comparison}
Y.~Gu \emph{et~al.}, ``Comparison between scalp eeg and behind-the-ear eeg for
  development of a wearable seizure detection system for patients with focal
  epilepsy,'' \emph{Sensors}, vol.~18, no.~1, p.~29, 2018.

\bibitem{do2014wireless}
B.~G. Do~Valle \emph{et~al.}, ``Wireless behind-the-ear eeg recording device
  with wireless interface to a mobile device (iphone/ipod touch),'' in
  \emph{2014 36th Annual International Conference of the IEEE Engineering in
  Medicine and Biology Society}.\hskip 1em plus 0.5em minus 0.4em\relax IEEE,
  2014, pp. 5952--5955.

\bibitem{9353647}
A.~Hamid \emph{et~al.}, ``The temple university artifact corpus: An annotated
  corpus of eeg artifacts,'' in \emph{2020 IEEE Signal Processing in Medicine
  and Biology Symposium (SPMB)}, 2020, pp. 1--4.

\bibitem{8768041}
C.-Y. Chang \emph{et~al.}, ``Evaluation of artifact subspace reconstruction for
  automatic artifact components removal in multi-channel eeg recordings,''
  \emph{IEEE Transactions on Biomedical Engineering}, vol.~67, no.~4, pp.
  1114--1121, 2020.

\bibitem{mognon2011adjust}
A.~Mognon \emph{et~al.}, ``Adjust: An automatic eeg artifact detector based on
  the joint use of spatial and temporal features,'' \emph{Psychophysiology},
  vol.~48, no.~2, pp. 229--240, 2011.

\bibitem{lawhern2012detection}
V.~Lawhern \emph{et~al.}, ``Detection and classification of subject-generated
  artifacts in eeg signals using autoregressive models,'' \emph{Journal of
  neuroscience methods}, vol. 208, no.~2, pp. 181--189, 2012.

\bibitem{8584791}
M.~Khatwani \emph{et~al.}, ``Energy efficient convolutional neural networks for
  eeg artifact detection,'' in \emph{2018 IEEE Biomedical Circuits and Systems
  Conference (BioCAS)}, 2018, pp. 1--4.

\bibitem{9137056}
Hasib-Al-Rashid \emph{et~al.}, ``A low-power lstm processor for multi-channel
  brain eeg artifact detection,'' in \emph{2020 21st International Symposium on
  Quality Electronic Design (ISQED)}, 2020, pp. 105--110.

\bibitem{roy2019machine}
S.~Roy, ``Machine learning for removing eeg artifacts: Setting the benchmark,''
  2019.

\bibitem{qendro2021high}
L.~Qendro \emph{et~al.}, ``High frequency eeg artifact detection with
  uncertainty via early exit paradigm,'' 2021.

\bibitem{Kim2021}
\BIBentryALTinterwordspacing
D.~K. Kim \emph{et~al.}, \emph{Fast Automatic Artifact Annotator for EEG
  Signals Using Deep Learning}.\hskip 1em plus 0.5em minus 0.4em\relax Cham:
  Springer International Publishing, 2021, pp. 195--221.
\BIBentrySTDinterwordspacing

\bibitem{9644949}
T.~M. Ingolfsson \emph{et~al.}, ``Towards long-term non-invasive monitoring for
  epilepsy via wearable eeg devices,'' in \emph{2021 IEEE Biomedical Circuits
  and Systems Conference (BioCAS)}, 2021, pp. 01--04.

\bibitem{pullini2019mr}
A.~Pullini \emph{et~al.}, ``Mr. wolf: An energy-precision scalable parallel
  ultra low power soc for iot edge processing,'' \emph{IEEE J. Solid-State
  Circuits}, vol.~54, no.~7, pp. 1970--1981, 2019.

\bibitem{kartsch2019biowolf}
V.~Kartsch \emph{et~al.}, ``Biowolf: A sub-10-mw 8-channel advanced
  brain--computer interface platform with a nine-core processor and ble
  connectivity,'' \emph{IEEE Trans. Biomed. Circuits Syst.}, vol.~13, pp.
  893--906, 2019.

\bibitem{9413712}
A.~Saeed \emph{et~al.}, ``Learning from heterogeneous eeg signals with
  differentiable channel reordering,'' in \emph{ICASSP 2021 - 2021 IEEE
  International Conference on Acoustics, Speech and Signal Processing
  (ICASSP)}, 2021, pp. 1255--1259.

\bibitem{ISLAM2016287}
\BIBentryALTinterwordspacing
M.~K. Islam \emph{et~al.}, ``Methods for artifact detection and removal from
  scalp eeg: A review,'' \emph{Neurophysiologie Clinique/Clinical
  Neurophysiology}, vol.~46, no.~4, pp. 287--305, 2016.
\BIBentrySTDinterwordspacing

\bibitem{mallat1989theory}
S.~G. Mallat, ``{A theory for multiresolution signal decomposition: the wavelet
  representation},'' \emph{IEEE Trans. Pattern Anal. Mach. Intell.}, vol.~11,
  no.~7, pp. 674--693, 1989.

\bibitem{benatti2014towards}
S.~Benatti \emph{et~al.}, ``{Towards EMG control interface for smart
  garments},'' in \emph{Proc. ACM ISWC}, 2014, pp. 163--170.

\bibitem{7833731}
------, ``Scalable eeg seizure detection on an ultra low power multi-core
  architecture,'' in \emph{2016 IEEE Biomedical Circuits and Systems Conference
  (BioCAS)}, 2016, pp. 86--89.

\bibitem{le2020scaling}
T.~T. Le \emph{et~al.}, ``Scaling tree-based automated machine learning to
  biomedical big data with a feature set selector,'' \emph{Bioinformatics},
  vol.~36, no.~1, pp. 250--256, 2020.

\bibitem{agtabular}
N.~Erickson \emph{et~al.}, ``Autogluon-tabular: Robust and accurate automl for
  structured data,'' \emph{arXiv preprint arXiv:2003.06505}, 2020.

\bibitem{H2OAutoML20}
\BIBentryALTinterwordspacing
E.~LeDell \emph{et~al.}, ``{H2O} {A}uto{ML}: Scalable automatic machine
  learning,'' \emph{7th ICML Workshop on Automated Machine Learning (AutoML)},
  July 2020.
\BIBentrySTDinterwordspacing

\bibitem{feurer-neurips15a}
M.~Feurer \emph{et~al.}, ``Efficient and robust automated machine learning,''
  in \emph{Advances in Neural Information Processing Systems 28 (2015)}, 2015,
  pp. 2962--2970.

\bibitem{Rokach2005}
\BIBentryALTinterwordspacing
L.~Rokach \emph{et~al.}, \emph{Decision Trees}.\hskip 1em plus 0.5em minus
  0.4em\relax Boston, MA: Springer US, 2005, pp. 165--192.
\BIBentrySTDinterwordspacing

\bibitem{Breiman1983ClassificationAR}
L.~Breiman \emph{et~al.}, ``Classification and regression trees,'' 1983.

\bibitem{8257019}
V.~Shah \emph{et~al.}, ``Optimizing channel selection for seizure detection,''
  in \emph{2017 IEEE Signal Processing in Medicine and Biology Symposium
  (SPMB)}, 2017, pp. 1--5.

\end{thebibliography}

\end{document}